\documentstyle[12pt]{article}
\newcommand{\beq}{\begin{equation}}
\newcommand{\eeq}{\end{equation}}
\newcommand{\ber}{\begin{eqnarray}}
\newcommand{\eer}{\end{eqnarray}}
\begin{document}  
\title{Factorization in Exclusive and Seminclusive Decays 
       and Effective Theories for Massless Particles}

\author{ 
{\it Ugo Aglietti, Guido Corb\`o}
 \\
$~~~$\\
Dipartimento di Fisica, Universit\` a di Roma \lq La Sapienza\rq \\ 
INFN, Sezione di Roma I, P.le A. Moro 2, 00185 Roma, Italy.
\\}
\date{}
\maketitle
\begin{abstract}
\noindent 

We prove rigorously factorization in the seminclusive decay
$B\rightarrow D^{(*)} + jet$ using the large energy effective theory.
It is also shown that this effective theory is unable to consistently 
describe completely exclusive processes, such as the decay
$B\rightarrow D^{(*)}+\pi(\rho)$, and therefore also 
related properties such as factorization. This is due to an
oversimplification of transverse momentum dynamics.
We present a variant of the large energy effective theory, i.e. a new
effective theory for massless particles, 
which properly takes into account transverse momentum dynamics
and is therefore the natural framework to study exclusive 
non-leptonic decays.
Finally, it is shown that the collinear instability of the large energy
effective theory disappears when seminclusive observables are considered.

\end{abstract}

\newpage

\section{Introduction and Summary of the Results}

The Large Energy Effective Theory $(LEET)$
has been introduced in ref.\cite{dgr} in
connection with factorization of exclusive non leptonic decay amplitudes
of heavy mesons, i.e. in processes like
\beq\label{nld}
B~\rightarrow~D^{(*)}+\pi(\rho).
\eeq
By factorization we mean that the decay (\ref{nld}) consists of
two independent, i.e. non-interacting, subprocesses:
the transition of a $B$ meson into a $D^{(*)}$ meson by the action
of the weak $b\rightarrow c$ current (carrying a momentum $q$),
\beq
B~\rightarrow~D^{(*)},
\eeq
and the creation of a light meson by the action on the
vacuum of the light $ud$ current (carrying a momentum $-q$,
with $q^2=m_{\pi}^2(m_{\rho}^2)$),
\beq
0~\rightarrow~\pi(\rho).
\eeq
The importance of factorization is that it allows the prediction
of the rate of (\ref{nld}) in terms of the rates of two
simpler (and well known) processes.
To compute the non-leptonic rate under factorization
assumption, we need only the form factors
entering the semileptonic decay
\beq\label{semil}
B~\rightarrow~D^{(*)}~+~e~+~\nu_e
\eeq
and the $\pi(\rho)$ decay constant, the former 
entering the experimentally well known leptonic decay
\beq\label{lept}
\pi~\rightarrow~\mu+\nu_{\mu}
\eeq
(for experimental tests see  \cite{esperimenti}).
Factorization essentially means that there is no interaction between
the system composed of the $B$ and $D^{(*)}$ mesons and
the system composed of the $\pi$ or the $\rho$.
In other words factorization means that the exchange
of hadrons between the two systems does not give a correction
to the rate.
In perturbation theory, the interaction between these two systems
is represented by gluons connecting the valence quarks of the
heavy mesons with the valence quarks of the light meson, and
factorization means that these gluons have no dynamical effect.
A formal condition for factorization to hold is that the
one-loop anomalous dimension of the four fermion operator $O$ 
inducing the decay
(\ref{nld}) equals the sum of the one-loop anomalous dimensions
of the current $J_h$ inducing the semileptonic 
decay (\ref{semil}) and of the current 
$J_l$ inducing the leptonic $\pi$-decay (\ref{lept}):
\beq
\gamma_O~=~\gamma_{J_h}+\gamma_{J_l}.
\eeq
This condition is, in general, not satisfied \cite{polwis}. 

Let us present the qualitative ideas about
factorization in the decay (\ref{nld}). The aim of this paper
is to give a theoretical support to these ideas, or to disprove them,
by means of effective theories. 
We start from the picture of the process given by Bjorken
\cite{bjk}. 
The light quark and the light antiquark forming the $\pi(\rho)$
are created by the weak $ud$ current 
at the decay time in the same spatial point in a color singlet state
(let us neglect for the moment the $ud$ color octet current,
i.e. let us neglect hard gluon effects setting $C_+,C_-=1,$ 
see sec.(\ref{secexcl})). 
This implies that there is no net color charge at the decay time.
As the light quarks fly away, they separate from each other generating a 
growing color dipole. If the quarks are emitted with a very
large energy $E$, color dipole growth will be slow because of time
dilation. More specifically, if
\beq
\tau\sim 10^{-23}sec 
\eeq
is the time for the light pair to hadronize
into a light meson at rest, the time
$t$ for the formation of the meson in the decay (\ref{nld}) is
\beq
t~=~\gamma_L~\tau 
\eeq
where $\gamma_L=E/\mu\gg 1$ is the Lorentz factor and $\mu$ is the pair
mass.
As a consequence, the color dipole is very small when the light pair is
inside the heavy meson, so the interaction between the two systems
is negligible in the infinite energy limit,
\beq
E~\rightarrow~\infty.
\eeq
This picture suggests factorization of the non-leptonic decay
amplitude (\ref{nld}). The argument is not rigorous in many respects. 
Criticism is related to general properties of quantum field theory
dynamics.
Particles cannot be localized inside a region smaller than their
Compton wavelength 
\beq
\lambda_C=1/m.
\eeq 
This length is of the order of the confinement  radius (or more)
$r_C\sim 1/\Lambda_{QCD}\sim 1fm$ for a light quark:
\beq
\lambda_C~\sim~r_{C}.
\eeq
That means that the light quarks are created at a relative distance $d$
of the order of the confinement radius,
\beq
d~\sim~r_{C}.
\eeq
The increase of the emission energy $E$ boosts the
longitudinal momenta of the quarks and consequently shrinks
the system in the longitudinal direction; 
it does not change the transverse momenta and 
the transverse dimension of the system:
\beq
d_L~\sim~\frac{r_{C}}{\gamma},~~~~~~
d_T~\sim~r_{C}.
\eeq 
Therefore the color dipole strength is not zero at the decay time: 
it is large, of order $g~r_{C}$, where $g$ is the quark color charge.
A large color dipole means strong color interactions with
the heavy system and an (expected) substantial violation of factorization.

The above argument can be rephrased in terms of the 
uncertainty principle.
For the dipole field strength to be zero and factorization
to hold, we require the transverse separation $b$ between the light
quarks, the so-called impact parameter, to be zero:
\beq
b~=~0.
\eeq
As a consequence, the average
relative transverse momentum squared of the light pair is infinite:
\beq
\langle q_T^2 \rangle~=~\infty.
\eeq
This implies that a bound state $\mid \psi\rangle$, which is a state
with a finite relative transverse momentum between the constituents, 
\beq
\langle q_T^2\rangle~\sim~\Lambda_{QCD}^2,
\eeq
has zero overlap with the state $\mid b=0\rangle$:
\beq
\langle \psi\mid b=0\rangle~=~0.
\eeq
To sum up:
the probability of hadronization into a given meson state by
a light quark pair generating a vanishing color dipole, is zero.
The chain of arguments is the following:
\beq
(factorization)~\Rightarrow~b=0~\Rightarrow~\langle q_T^2 \rangle
=\infty~
\Rightarrow~\langle\psi\mid b=0 \rangle=0.~~~
\eeq
Furthermore, it is well known that if one localizes a particle in a region
smaller than $\lambda_C$, additional particle-antiparticle
pairs are created. The result is a mutiparticle state. 
The probability for a multiparticle state to hadronize into a single
meson is very small.

In conclusion, we believe that factorization cannot be rigorously proved
for the completely exclusive process (\ref{nld}).
The point to keep in mind is that factorization holds as long as 
we consider states of the light quark system with small impact
parameter
\beq
b~\sim~0.
\eeq
We can change our viewpoint and consider the seminclusive process
\beq\label{semincl}
B~\rightarrow~D^{(*)}+jet.
\eeq
The criticism above does not apply to this seminclusive process, 
in which the 
light quarks produce a jet of hadrons, instead of a single meson.
Squeezing the color dipole (necessary for factorization!), 
as we have seen, induces pair creation,
but this is not a problem for a jet, which is a multiparticle
state.  
The large relative transverse momentum squared of the squeezed
pair is not even a problem for the seminclusive decay.
Assume that the light quarks are created at a very small transverse
separation $b$, 
\beq
b~\ll~r_C.
\eeq
Because of the uncertainty principle, this implies a
broad range of relative transverse momenta,
\beq
\langle q_T^2 \rangle~\sim~\frac{1}{b^2}~\gg~\Lambda_{QCD}^2.
\eeq
Partons with high energy, producing jets, 
can sustain large transverse momentum fluctuations.
Their motion is specified by a space direction, so if we call
$\theta$ the deflection angle due to a transverse momentum
transfer $q_T$, we have:
\beq
\theta~\sim~\frac{q_T}{E}~\rightarrow~0
\eeq
in the `hard' limit
\beq
E~\rightarrow\infty,
\eeq
as long as we keep $b>0$, i.e. small but not zero.
Formally, we take the limits in the following order:
\begin{enumerate}
\item[$(i)$] we send $E\rightarrow\infty$ keeping $b$ constant but
not zero;
\item[$(ii)$] we send $b\rightarrow 0$ to have complete color screening
              and factorization.
\end{enumerate}
In words: the motion of the hard parton is not modified by 
transverse momentum fluctuations. These fluctuations
affect only parton shower development and
the hadronization process, i.e. the latter stages
of jet dynamics. In the seminclusive process we get rid of these
effects summing over all possible jet developments.

To summarize, we argue that factorization cannot be proved
rigorously in the exclusive process (\ref{nld}) due to
basic quantum field theory phenomena: these problems can be 
circumvented considering in place of the exclusive process
the seminclusive one (\ref{semincl}).

The rest of this paper is a formalization of the above ideas and
considerations. We will see that the intuitive idea that
a vanishing color dipole has no interactions, so its dynamics
is factorized, can be 
rigorously formalized in gauge theories through the use of Wilson loops.

Let us discuss now the organization of the paper and outline
the main results.
In sec.\ref{secleet} we review the basic elements of the $LEET$.
The physical content is very simple.
This effective theory describes massless quarks with a 
very large energy $E$ or, equivalently, quarks suffering soft 
interactions only. If $k$ is the momentum exchanged by the quark,
we require:
\beq
\mid k_{\mu}\mid~\ll~E.
\eeq 
This means that the motion of the effective quark
can be approximated by a straight line, with the velocity of light,
in the (soft) collisions. In other words, we consider the action
of the hard parton on soft quarks and gluons, but we neglect the
reaction of the soft particles on the hard parton (recoil effect).
This is very well seen looking at the propagator in configuration
space in an arbitrary external gauge field,
\beq
iS_F(x,0)~=~\int_0^{\infty} d\tau~\delta^{(4)}(x-n\tau)~P
e^{ i\int d^4y~J_{\mu}(y)~A^{\mu}(y) }
\eeq
where $J_{\mu}(y)$ is a color current associated with the classical
motion of the quark:
\beq
J_{\mu}(y)~=~g\int_0^{\tau} ds~ 
n_{\mu}~\delta^{(4)}(y-n s).
\eeq 
The propagator has support (i.e. it is not zero)
only on the points of the classical trajectory 
\beq
x_{\mu}~=~n_{\mu}\tau,~~~\tau~\geq~0,
\eeq 
and the interaction produces only a phase factor in color space
(Wilson P-line). The probability for the effective quark to reach
a given point is not modified by the interaction.
By this fact we mean that the reaction of the light particles on the 
effective quark is neglected.
All this is fine.
However, it has been shown in ref.\cite{agl} that the $LEET$
suffers from consistency problems.
We believe that the proof of the consistency of the $LEET$
is a preliminary step for any phenomenological application,
such as the justification of factorization.
The problem originates from the fact that the $LEET$
contains states with negative energies,
because the dispersion relation is of the form
\beq\label{disprel}
\epsilon~=~k_z,
\eeq
i.e. the modulus of the space momentum is missing
(we have taken for simplicity a motion along the $z$ axis).
The negative energies cannot be removed with a Lorentz
transformation as in the case of the $HQET$,
and render the model unstable. 
The effective quark $Q$ can decay into states with arbitrarily
negative energies emitting collinear particles:
\beq
Q~\rightarrow~Q~+~(coll.~part).
\eeq 
Therefore one-particle states are unstable, with the consequence
for example that: 
\beq
\mid Q\rangle_{in}~\neq~\mid Q\rangle_{out}.
\eeq
In sec.\ref{secinst} we review
the collinear instability phenomenon and we prove 
that it disappears when semi-inclusive observables are considered.
The idea is that of replacing the single particle states of the
effective quark $Q$ with cones centered along the line of motion
of $Q$ containing any particle. 

In sec.\ref{secexcl} we discuss the application 
of the $LEET$ to exclusive processes, and
in particular to the factorization of non-leptonic amplitudes.
As it is clear from the physical considerations done before,
the $LEET$ cannot describe any bound state
effect. It is clearly impossible for an effective quark and an effective 
antiquark, each one moving with uniform rectilinear motion, 
\ber
x_{\mu}^{(Q)}&=&n\tau,
\nonumber\\
x_{\mu}^{(\overline{Q})}&=&n\tau',
\eer
to produce any bound state. 
Technically, we will see that all the correlation functions describing
exclusive dynamics turn out to be singular and trivial in the $LEET$.
All this comes from neglecting transverse momentum dynamics
(just look at the dispersion relation (\ref{disprel})):
the effective quarks are created and remain at any time
at a fixed transverse separation, $b=0$ in the case of the local 
$ud$ current. 

In sec.\ref{secfac} we simply reinterpret the results of the
previous section. The singularity of the correlators related
to exclusive dynamics disappears once seminclusive processes
are considered, and their triviality does mean factorization
in this case.
We also discuss the connection of the $LEET$ with the
theory of color coherence, describing interjet activity 
in the perturbative $QCD$ framework \cite{dks}.

Having discovered that the $LEET$ is not the relevant
theory for studying
exclusive processes like (\ref{nld}), we may ask whether it
is possible to formulate another effective theory for
massless particles which does the job.
In sec.\ref{secnew} we introduce such a new effective theory 
which we call Modified Large Energy Effective Theory 
($\overline{LEET}$).
The idea is that of including in the dispersion relation (\ref{disprel})
the first correction dependent on the transverse momentum,
so that the latter is replaced by: 
\beq\label{ktdisp}
\epsilon~=~k_z+\frac{\vec{k}_T^2}{2E},
\eeq
where $\vec{k}_T^2=k_x^2+k_y^2$.
We will see that the effect of the transverse momentum in the 
propagator in configuration space is a diffusive one.
The propagator is not concentrated only on the classical trajectory,
as is the case for the $LEET$,
but it is diffused in a region which is growing with time.
The probability for the quark to remain into the classical
trajectory decays as the inverse of the time squared:
\beq
P(classical~path;~t)~\sim~\frac{1}{t^2}.
\eeq
The inclusion of transverse momentum makes the composite 
correlation functions non singular and not trivial at the same 
time. 
In general, the dynamics of the $\overline{LEET}$ is much 
richer and complicated
than that of the $LEET$, but it is still much
simpler than in the original Dirac theory. The simplification 
essentially occurs because pair creation is absent, so the
vacuum in the matter sector is trivial.
The $\overline{LEET}$ is not related to the theory of Wilson loops,
as it was the case of the $LEET$; as a consequence, for
example, the propagator  of the $\overline{LEET}$
cannot be computed in closed form in the interacting case,
as it was for the $LEET$.
The $\overline{LEET}$ bears some resemblance with the 
Non-Relativistic $QCD$ ($NRQCD$) \cite{lepage}, which
is an effective field theory with the non-relativistic
energy-momentum relation:
\beq\label{nonrel}
\epsilon~=~\frac{\vec{k}^2}{2m}
\eeq
where $m$ is the heavy quark mass (compare eq.(\ref{ktdisp}) with 
eq.(\ref{nonrel})).

In general, we believe that the $\overline{LEET}$ is the 
effective theory for massless particles which is the right framework
to study exclusive processes, and consequently also factorization
of exclusive non-leptonic decays.

Sect.\ref{secconcl} contains the conclusions of our 
investigation.

\section{The Large Energy Effective Theory}
\label{secleet}

Let us briefly review the basic elements of the LEET \cite{dgr,agl}.
It is derived decomposing the momentum $P$ of a massless quark 
into a `classical part' $En$ and a fluctuation $k$:
\beq
P~=~En+k.
\eeq
$n$ is a light-like vector $n^2=0$, normalized by the condition
$v\cdot n=1$, where $v$ is a reference time-like vector, $v^2=1,~v_0>0$.
Therefore $E$ is the classical quark energy in the rest frame of $v$ and
has to
be considered large,
\beq
\mid k_{\mu} \mid~\ll~E.
\eeq
The propagator is given by
\ber\label{longway}
&&iS_F(En+iD)~=~\frac{i}{E\hat{n}+i\hat{D}+i\epsilon}
\nonumber\\
&=&(E\hat{n}+i\hat{D})\frac{i}{2En\cdot i D -D^2
+\sigma_{\mu\nu}gG^{\mu\nu}/2+i\epsilon}
\nonumber\\
&=&\frac{ \hat{n} }{2}\frac{i}{n\cdot iD+i\epsilon}~+~
\frac{1}{2E}~\frac{ \hat{n} }{2}\frac{i}{n\cdot iD+i\epsilon}
(-i)D^2~\frac{i}{n\cdot iD+i\epsilon}
\nonumber\\
&+&\frac{ig}{4E}\frac{ \hat{n} }{2}\frac{i}{n\cdot iD+i\epsilon}~
\sigma_{\mu\nu}G^{\mu\nu}~\frac{i}{n\cdot iD+i\epsilon}
+\frac{i\hat{D}}{2E}~\frac{i}{n\cdot iD+i\epsilon}
+O\left(\frac{1}{E^2}\right).~~~~~~
\eer
where $\hat{a}=\gamma_{\mu}a^{\mu}$, 
$\sigma_{\mu\nu}=i[\gamma_{\mu},\gamma_{\nu}]/2$,
$D_{\mu}=\partial_{\mu}-igA_{\mu}$ and 
$-igG_{\mu\nu}=[D_{\mu},D_{\nu}]$. 
We made the replacement $k\rightarrow iD$ to include
magnetic interactions (we assume the usual convention
that $\psi(x)\sim \exp(-ik\cdot x)$).

We treat $1/E$ corrections as perturbations to the leading
term in eq.(\ref{longway}), so that the $LEET$ propagator is
\beq
iS_F(k)~=~\frac{\hat{n}}{2}~\frac{i}{n\cdot k+i\epsilon}
\eeq
The energy-momentum relation (the pole of the propagator) is
\beq
\epsilon~=~\vec{u}\cdot \vec{k}
\eeq
where $\vec{u}=\vec{n}/n_0$ is the kinematical velocity, $\mid\vec{u}\mid=1$.

\noindent
The propagator in configuration space is the scalar density
of a particle moving along a ray with the velocity of light: 
\beq
iS_F^0(x)~=~\frac{ \hat{n} }{2}~\int_0^{\infty}
d\tau~\delta^{(4)}(x-n\tau)~
=~\frac{\hat{n}}{2}~\theta(t)~\frac{\delta^{(3)}(\vec{x}-\vec{u}t)}{n_0}
\eeq
The interaction with the gauge field produces a P-line factor
joining the origin with the point $x$ along the light-like
trajectory specified by $n$:
\beq
iS_F(x)~=~iS_F^0(x)~P\exp\Bigg[~ig\int_0^{t/n_0}dt'~n_{\mu}A^{\mu}(nt')~\Bigg].
\label{prop}\eeq
Note the factorization of both the spin and the color degrees of
freedom.
The propagator can be written in a 
completely covariant form, reminiscent of the
Dyson formula for the $S$-matrix:
\beq
iS_F(x)~=~\frac{ \hat{n} }{2}~\int_0^{\infty} d\tau~\delta^{(4)}(x-n\tau)
{\rm P}\exp\Bigg[~i\int d^4 y~J_{\mu}(y)~A^{\mu}(y) ~\Bigg]
\eeq
where $J_{\mu}$ is a classical current
\beq
J_{\mu}(y)~=~g\int_0^{\tau}d s~n_{\mu}~\delta^{(4)}(y-n s).
\eeq
In momentum space, we have:
\beq\label{puntini}
iS_F(k)~=~\frac{ \hat{n} }{2}~\int_0^{\infty} d\tau~
e^{i(k\cdot n+i\epsilon)\tau}~
\Bigg[1+ig\int \frac{d^4p}{(2\pi)^4}~n_{\mu}A^{\mu}(-p)~  
\frac{e^{i(p\cdot n+i\epsilon)\tau}-1}{i(p\cdot n+i\epsilon)}+\ldots\Bigg]
\eeq
The Feynman rule for the vertex is
\footnote{
This formula is easily derived from eq.(\ref{puntini}) making use of the
eikonal identity
$$
\frac{1}{k\cdot n+i\epsilon}-
\frac{1}{(k+p)\cdot n+i\epsilon}~=~
\frac{n\cdot p}{(n\cdot k+i\epsilon)((k+p)\cdot n+i\epsilon)}.
$$}
\beq
V~=~ign_{\mu}t^a.
\eeq
The $LEET$ Lagrangian, omitting the spin structure, is
\beq
{\cal L}(x)~=~
\overline{Q}_n(x)~in\cdot D~Q_n(x)~+~O\Big(\frac{1}{E}\Big).
\eeq
The $LEET$ describes massless particles suffering soft interactions only.

\section{The Instability and Its Cure}
\label{secinst}

Let us briefly review the instability phenomenon \cite{agl}.
A simple example is the collision between an effective quark $Q$ with 
(residual) momentum $k$ and a full quark $q$ with momentum $p$.
Energy and momentum conservation give
\ber
\mid\vec{p}\mid + \vec{u}\cdot\vec{k}&=& \mid\vec{p'}\mid
+\vec{u}\cdot\vec{k'}
\nonumber\\
\vec{p}+\vec{k}&=&\vec{p'}+\vec{k}'
\eer
where $q$ is taken massless for simplicity.
Let us assume that $p+k$ is a time-like vector, $(p+k)^2>0$.
In the COM frame, $\vec{p}+\vec{k}=0$,
with $\vec{n}$ oriented along the $+z$ axis, we have
\beq
     p  (1-\cos\theta)~=~p' (1-\cos\theta')~=~\cal{E}
\eeq
where $\theta$ is the angle between the quark 3-momentum and $\vec{n}$.
The prime denotes final state quantities and
\beq
\cal{E}~=~\sqrt{(p+k)^2}~\ll~E
\eeq 
is the total energy of the system in the $LEET$;
it is expected to be of the order $\Lambda_{QCD}$ in $QCD$ applications,
while $E$ is the hard scale of the process.
The instability originates because
\beq\label{diverg}
p'~=~\frac{\cal{E}}{1-\cos\theta'}~\rightarrow~\infty
\eeq
when
\beq
\theta'~\rightarrow~0.
\eeq
It is related to the emission of particles in the forward direction
$\vec{n}$, the flight direction of $Q$.

Let us assume now a finite angular resolution $\delta>0$ of the
detectors. This implies that $Q$ cannot be distinguished from
almost-collinear partons. We consider a cone of half-opening angle 
$\delta$ with the axis along $\vec{n}$. 
We have that $q$ is observed as an individual particle
if it is emitted in the final state outside the cone,
\beq
\theta'~>~\delta.
\eeq 
In this case the energy is bounded by
\beq
p'~<~ \frac{\cal{E}}{1-\cos\delta}~<~\infty,
\label{bound}\eeq
and cannot diverge anymore. Therefore,
the divergence (\ref{diverg}) does not occur for an observable parton.
On the other hand, if the final parton is inside the cone,
\beq
\theta'~<~\delta,
\eeq
the finite angular resolution makes impossible to 
detect $q$ and $Q$ as separated particles and to measure
their individual energies. A single particle (jet) is observed
with the sum of the parton energies
\beq
p'+\epsilon'~=~\frac{\cal{E}}{1-\cos\theta'}
-\cos\theta'\frac{\cal{E}}{1-\cos\theta'}~=~\cal{E}~<~\infty
\eeq
The individual energies are separately divergent but the sum
is finite (small) by assumption (it equals the initial energy). 
The instability is therefore eliminated by the angular separation
requirement. 
We stress that the cone has to be centered around the vector $\vec{n}$
and not around the residual momentum $\vec{k}$. 
In a general process, the instability is eliminated
replacing the states of the effective
quarks with cones centered around their line of flight.

\subsection{Matching}

The angular resolution $\delta$ is a parameter generated {\it inside} 
the effective theory by a consistency requirement, but its 
value is completely arbitrary.
Let us see now the restrictions imposed on the values on $\delta$
by comparison with the full theory, the so-called matching.
Before the expansion, the energy-momentum relation is
\ber\label{drl}
\epsilon&=&\sqrt{ P_Z^2+\vec{P_T}^2 }-E~=~
\sqrt{ (E+k_Z)^2+\vec{k_T}^2 }-E
\nonumber\\
&=&\mid E+k_Z\mid -E+\frac{\vec{k}_T^2}{2\mid E+k_Z\mid}-
\frac{\vec{k}_T^4}{8\mid E+k_Z\mid^3}+O(\vec{k}_T^6)
\eer 
where in the last line a power expansion in $k_T$ has been done.

Going to the $LEET$ implies two different kinds of approximations:
\begin{enumerate}
\item[$(i)$]
Neglecting the modulus in eq.(\ref{drl}). 
Keeping the modulus of the momentum would imply non-local operators
in the effective theory, which are against the spirit of
the effective theories themselves. Omitting the modulus
we loose the positivity of the energy and the 
instability problem comes up. As we have seen, this problem is
solved redefining the observables, i.e. considering seminclusive ones.
This approximation is unavoidable.
\item[$(ii)$]
Neglecting the transverse momentum $k_T$ in eq.(\ref{drl}).
The effect of this approximation is to mistreat bound state
dynamics. As we will see, the correlators describing bound states,
come out trivial and singular at the same time, as a consequence
of this approximation.
It is possible to remedy to this problem keeping transverse
momentum terms in eq.(\ref{drl}):
\beq\label{drl2}
\epsilon~=~k_Z+\frac{k_T^2}{2E}~-\frac{k_Z~k_T^2}{2E^2}~
+\frac{k_Z^2~k_T^2}{2E^3}-\frac{k_T^4}{8E^3}+\ldots
\eeq
Note, in particular, that keeping terms in the residual
momentum to any finite order (i.e. $1/E$ corrections) does not stabilize
the theory, as one could instead naively think.
\end{enumerate}
The $LEET$ dispersion relation is therefore:
\beq
\epsilon~=~k_Z.
\eeq
Let us take $k_T=0$ in eq.(\ref{drl}).
We see that the $LEET$ is a good approximation as long as
\beq
\mid k_Z \mid ~\leq~ E
\eeq
The above condition for the final state momentum, 
$\mid k_Z'\mid\leq E$, implies
\beq
\frac{\cos\theta'}{1-\cos\theta'}~\leq~\frac{E}{\cal{E}}.
\eeq
Comparing with eq.(\ref{bound}) we have, for small angles:
\beq\label{match}
\delta~\geq~\delta_C~=~\sqrt{ \frac{2 \cal{E}}{E} }~~~~~~~~~~~~~~~~~~
(\delta~\ll~1)
\eeq
Eq.(\ref{match}) is our final result. It says that the LEET is a good
approximation of the full theory as long as $\delta$ is
taken above a critical value $\delta_C$ below which fictious energies
come into the game.

The angular resolution $\delta$ we have introduced is similar
in spirit to the parameter introduced to cancel collinear singularities 
in massless theories \cite{sterwein}. There are however differences.
Collinear singularities induce a divergence in the cross section,
while in our case cross sections are finite but involve states
with infinite energies. Furthermore, in the process 
we considered, Rutherford scattering, there are 
no collinear singularities. 

\section{Exclusive Processes and Factorization}
\label{secexcl}

In this section we show that the $LEET$ cannot be used 
consistently to describe
exclusive processes such as the decay (\ref{nld}). 
The staring point is the reduction formula, according to which
any scattering matrix element $S_{fi}$ can be derived from the
correlation functions of the theory \cite{symanzik},
\beq
G(x_1,x_2,\ldots,x_n)~=~
\langle 0\mid T \phi(x_1)\phi(x_2)\ldots\phi(x_n)\mid 0\rangle
\eeq
where $\phi$ denotes generically a field, and $T$ is the
Dyson time-ordering operator.
This assumption implies that if a given process cannot be derived 
from any correlation function of the theory, the latter does
not describe the process.

\subsection{Spectroscopy}

Let us begin by considering the simplest correlator, the propagator of a
light meson such as a $\pi$ or a $\rho$, i.e. the 2-point function
\beq\label{dibase}
C(x)~=~
\langle 0\mid T O(\vec{x},t)O^{\dagger}(\vec{0},0)\mid 0\rangle
\eeq
where
\beq
O(x)~=~\overline{u}(x)\Gamma d(x).
\eeq 
and $\Gamma$ is a matrix in the Dirac algebra.
We can set to zero the residual spatial momentum of the meson
because we are interested in internal dynamics only, so we consider
\beq\label{settozero}
C(t)~=~\int d^3x~C(t,\vec{x}).
\eeq
The correlation $C(t)$ has the spectral decomposition 
\cite{symanzik2} for $t>0$ 
\ber\label{spectral}
C(t)&=&\sum_n~ 
\frac{ \mid \langle 0\mid O(0)\mid M_n(\vec{p}=0)\rangle\mid^2 }{2m_n}~ 
      e^{-im_nt} 
\nonumber\\
&+& (multiparticle~states)
\eer
where $M_n$ denotes all the meson states and
the lightest state, depending on the spin-parity of the current,
can be a $\pi$, a $\rho$, etc.
\footnote{
The states have been normalized in a covariant way:
$$
\langle B(p)\mid B(p')\rangle~=~(2\pi)^3~ 
2E(p)~\delta^{(3)}~(\vec{p}-\vec{p}').
$$ }.
$C(t)$ has the functional integral representation:
\beq
C(t)~=~\int d^3 x~N~\int [dA_{\mu}]~e^{iS_{eff}[A_{\mu}]}~
\langle 0\mid T O(\vec{x},t)O^{\dagger}(\vec{0},0)\mid 0\rangle_{A}
\eeq
where 
\beq
N^{-1}~=~\int [dA_{\mu}]~e^{iS_{eff}[A]} 
\eeq 
is a normalization factor,
$S_{eff}[A]=S_{YM}[A]-n_l/2\log\det[i\hat{D}]$ is the effective 
gauge field action, and $n_l$ is the number of light flavors.
The $T$-ordered product with the subscript $A$ means the 
expectation value in the fermionic vacuum in 
a background gauge field $A_{\mu}$.
The Wick contraction gives:
\beq\label{ricorre}
\langle 0\mid T O(\vec{x},t)O^{\dagger}(\vec{0},0)\mid 0\rangle_{A}~
=~-Tr[~iS_F(x\mid 0;A_{\mu})\tilde{\Gamma}~iS_F(0\mid x;A_{\mu})\Gamma~]
\eeq
where $\tilde{\Gamma}=\gamma_0\Gamma^{\dagger}\gamma_0$.
Substituting the effective quark propagator, eq.(\ref{prop}), 
for $iS_F(x\mid 0)$, 
and an analogous effective propagator for the antiquark, 
\beq
iS_F(0\mid x)~\rightarrow~iS_{eff}^{\overline{Q}}(0\mid x)~
=~\frac{\hat{n}}{2}~\theta(t)~\frac{\delta^{(3)}(\vec{x}-\vec{u}t)}{n_0}~
P~e^{ i\int_x^0 A_{\mu}~dx^{\mu} },
\eeq
we derive:
\ber
&&C(t)=
\nonumber\\
&=&-\theta(t)^2\int d^3 x
\frac{\delta^{(3)}(\vec{x}-\vec{u}t)^2}{n_0^2}~
Tr\Bigg[\frac{\hat{n}}{2}\tilde{\Gamma}\frac{\hat{n}}{2}\Gamma\Bigg]~
Tr\Bigg[ Pe^{ig\int_0^x A_{\mu}~dx^{\mu}} I
         Pe^{ig\int_x^o A_{\mu}~dx^{\mu}} I\Bigg]
\nonumber\\
&=&-\frac{\theta(t)}{n_0^2}~
Tr_{spin}\Bigg[\frac{\hat{n}}{2}\tilde{\Gamma}
\frac{\hat{n}}{2}\Gamma\Bigg]~Tr_{col}\Big[~I~\Big]~
\delta^{(3)}_{\Lambda}(\vec{x}=0)
\eer
where $Tr_{col}[I]=N_C=3$. A few remarks are in order.
\begin{itemize}
\item[$(i)$]
There is an ultraviolet ($UV$) power divergence coming from the product
of the two
delta functions, which we have regularized with a sharp cut-off on the
spatial momenta
\beq
\delta^{(3)}_{\Lambda}(\vec{x})~=~\int^{\Lambda} \frac{d^3 k}{(2\pi)^3}~
e^{i\vec{k}\cdot\vec{x}}~
=~\frac{\Lambda^3}{6\pi^2}~\delta_{\vec{x},0}
\eeq
This divergence originates from the fact that the $LEET$ quarks
are created in the same point by the (local) current
and move along the same (classical) trajectory.
\item[$(ii)$]
The P-lines of the quark and the antiquark are one the inverse of the
other, so they give the unit operator in color space, 
independent of the gauge field $A_{\mu}$.
The functional integration over $A_{\mu}$ is therefore trivial. 
This implies that there is not any
color interaction, i.e. there are no gluons exchanged between the quarks: 
the theory is free.
\end{itemize}
The main point is that
the correlator is constant in time: comparing with the spectral
decomposition (\ref{spectral})
we derive that all the meson masses vanish:
\beq
m_n~=~0~~~~~~~~~any~n.
\eeq
Spectroscopy disappeared, 
but this is an infrared property
which should be represented by an effective theory.  
As we will see in sec.\ref{secnew}, it is still possible to 
build up an effective theory for massless particles in which 
spectroscopy is not made trivial.

Furthermore, since the contribution of any state to the correlator is 
time-independent, there is no way to separate 
{\it in principle} the contribution of
any selected meson (i.e. any exclusive state), such as a $\pi$ or a
$\rho$. 
The simultaneous singular behaviour and, more important, trivial
behaviour of the effective
correlator is related to neglecting altogether transverse
momentum dynamics (see sec.\ref{secnew}).

We conclude that the correlator (\ref{dibase}) describes the propagation
of a {\it jet} in a color singlet state, i.e. a state defined in a
seminclusive way, in which it is not possible to separate any 
exclusive state.

\subsection{Nonleptonic Decays}

Let us consider now the non-leptonic decay (\ref{nld}).
The rate can be computed, according to the standard reduction
formula, from the values of the following 4-point correlation
function
\beq\label{c4}
C_4(z,y,x)~=~\langle 0\mid T~O(z)~O_D(y)~\cal{H}_{W}(0)~O_B^{\dagger}(x)
\mid 0\rangle
\eeq
taking the asymptotic limits
\beq
t_x~\rightarrow~-\infty,~~~~~ 
t_y~\rightarrow~+\infty,~~~~~
t_z~\rightarrow~+\infty. 
\eeq
$O_B=\overline{q}\Gamma b$ and $O_D=\overline{q}\Gamma c$ are two
interpolating fields for the $B$ and $D^{(*)}$ mesons respectively.
$\cal{H}_W$ is the effective non-leptonic weak hamiltonian, which
may be written as
\beq\label{effham}
\cal{H}_W(x)~=~\frac{G_F}{\sqrt{2}}~\Big[C_1 O_1(x)~+~C_8
O_8(x)\Big]
\eeq
where $G_F$ is the Fermi constant and $O_1(x)$ and $O_8(x)$ are
local 4-fermion operators which are the product of two 
singlet and octet currents in color space, 
\beq
O_i(x)~=~\overline{d}(x)~\gamma_{\mu L}\xi_i~u(x)~
\overline{c}(x)~\gamma^{\mu}_L\xi_i~b(x)
\eeq
where $\gamma_{L}^{\mu}=\gamma^{\mu}(1-\gamma_5)$,
$\xi_i=1,t^a$ for $i=1,8$ respectively, 
and $C_1$ and $C_8$ are Wilson coefficients resumming hard
gluon effects of the form $\alpha_S^n\log^k(m_W^2/m_b^2),$
$k\leq n$ \cite{amgl}.

Inserting the expression (\ref{effham}) for the weak hamiltonian
in eq.(\ref{c4}), and
making the Wick contractions as we did for the 2-point function, 
we derive:
\ber\label{factoriz}
&&~~~~\langle 0\mid T~O(z)~O_D(y)~\cal{H}_{W}(0)~O_B^{\dagger}(x)
\mid 0\rangle_{A}
\nonumber\\
&&=\sum_{i=1,8}~C_i~
Tr\Big[~iS_c(y\mid 0;A_{\mu})~\gamma_{\mu L}\xi_i~iS_b(0\mid x;A_{\mu})~
\tilde{\Gamma}_B~iS_{s}(x\mid y;A_{\mu})~\Gamma_D~\Big]
\nonumber\\
&&~~~~~~~~~~~Tr\Big[~iS_d(z\mid 0;A_{\mu})~\gamma^{\mu}_L\xi_i~
iS_u(0\mid z;A_{\mu})~\Gamma~\Big],
\eer
where we have taken
a valence strange quark $(q=s)$ to avoid unimportant contractions.
Each contribution involves the product
of two separate fermionic traces.
The term with $i=8$ (i.e. the contribution of the
octet operator $O_8$) does not contribute to the
correlation because the trace of the light pair in color space
involves a matrix $t^a$
and is therefore proportional to
\beq
Tr\Bigg[ P~e^{i\int_0^z A_{\mu}~dx^{\mu}}~t_a~
         P~e^{i\int_z^o A_{\mu}~dx^{\mu}}~I~\Bigg]~
=~Tr~t^a~=~0.
\eeq
The physical reason of this result is that a quark and an antiquark 
in the same spatial point in a color
singlet state cannot emit a gluon and go into an octet state:
the dipole field strength is zero.

The only contribution comes from the term with $i=1$ (i.e. from $O_1$
only). 
The correlation (\ref{factoriz}) has a factor describing $ud$ dynamics
similar to the 2-point function (\ref{ricorre}),
\ber\label{matrcor}
\langle 0\mid T O(z) J_{\mu}^{u\rightarrow d}(0)\mid 0\rangle
&=&-{\rm Tr}\Big[~iS_d(z\mid 0)~\gamma^{\mu}_L~iS_u(0\mid z)~\Gamma~\Big]
\\
&=&-\frac{\theta(t)}{n_0^2}~
Tr_{spin}\Bigg[\frac{\hat{n}}{2}~\Gamma~
\frac{\hat{n}}{2}~\gamma^{\mu}_L\Bigg]~Tr_{col}\Big[~I~\Big]~
\delta^{(3)}_{\Lambda}(\vec{x}=0)
\nonumber\eer
where $J_{\mu}(x)^{u\rightarrow d}=\overline{d}(x)\gamma_{\mu}^L u(x)$.
This implies similar considerations as those given for the 2-point
function:
the dynamics of the light pair is independent of the gauge field,
so the $u$ and $d$ quarks do not have any interaction
with the heavy system (the $B$ and $D^{(*)}$ mesons)
and do not have any reciprocal interactions either.
Therefore, there is factorization in the sense that 
the $u-d$ system does not interact with the heavy system,
but there is not any interaction either between the quarks 
giving rise to the $\pi$ or the $\rho$ meson.
A given exclusive channel such as (\ref{nld})
cannot be selected {\it in principle} from the correlation.

To summarize, we say that 
the state created by the $ud$ current acting on the vacuum is a jet,
an indistinguishable ensemble of exclusive states  
(consisting of strictly collinear $u$ and $d$ quarks in lowest order).
The $LEET$ is not adequate to discuss issues
such as factorization in exclusive channels. 

\subsection{Perturbative Expansion}
\label{secexclpert}

If we compute the meson propagator (\ref{dibase}) 
in the full theory (in ordinary 4-momentum space), 
we find in lowest oder perturbation theory
\footnote{We neglect the numerator structure of the amplitude
because we are looking only at infrared singularities,
which appear as zeros of the denominator.}:
\ber\label{nonapprox}
 C_F &\sim&\int d^4
k~\frac{1}{(xP+k)^2+i\epsilon}~\frac{1}{((1-x)P-k)^2+i\epsilon}
\nonumber\\
&\sim &
\int d^4k
\frac{1}{k_0-k_z+k^2/(2xE)+i\epsilon}
\frac{1}{k_0-k_z-k^2/(2(1-x)E)-i\epsilon}
\nonumber\\
&\sim &\int \frac{d^2 k_T ~ d k_+ ~ d k_-}
{(k_-+k_+k_-/(2xE)-k_T^2/(2xE)+i\epsilon)}
\nonumber\\
&&~~~\frac{1}{(k_--k_+k_-/(2(1-x)E)+k_T^2/(2(1-x)E)-i\epsilon)}
\nonumber\eer
where we have taken an external momentum $P=En$
with $n=(1;0,0,1)$ and $E>0$. 
The variable $x$ represents the quark momentum fraction in the
infinite momentum frame ($0\leq x \leq 1$)
\footnote{We assume that the quark momentum 
distribution in the meson $q(x)$ is not singularly peaked at the endpoints 
$x=0,1$, so that $xE$ and $(1-x)E$ can always be considered
as large energies. A physical justification of this assumption
comes from an expected Sudakov suppression of the elastic region.}.
$k_+=k_0+k_z$ and $k_-=k_0-k_z$ are the usual light-cone variables
and $k_T^2=\vec{k}_T^2$.
The poles in the $k_-$-plane are located at
\beq
k_-~=~\frac{k_T^2/(2xE)-i\epsilon}{1+k_+/(2xE)},~~~~~
k_-~=~\frac{-k_T^2/(2(1-x)E)+i\epsilon}{1-k_+/(2(1-x)E)}
\eeq
Assuming a cutoff $\Lambda$ on $k_+$ satisfying $\Lambda\ll E$, the
integral is approximated by
\beq\label{approx}
C_F~\sim~\int d^2 k_T~\int d k_+ \int d k_-~
\frac{1}{k_- - k_T^2/(2xE)+i\epsilon}~
\frac{1}{k_-+k_T^2/(2(1-x)E)-i\epsilon}
\eeq
There is a pinching of the poles in the $k_-$-plane for $k_T=0$:
in other words, the integration contour is trapped 
between two poles which coalesce in the limit $k_T\rightarrow 0$.
The integral is logarithmically divergent 
\footnote{The infrared logarithmic singularity originates because the integral
(\ref{nonapprox}) does not contain any scale ($P^2=0$), so it is of the
form $\int
d^4k/(k^2)^2$, as can be seen explicitly introducing a Feynman
parameter.} with $\epsilon$:
\beq
C_F~\sim~\int \frac{dk_T^2}{k_T^2-i\epsilon}~\sim~\log\frac{1}{\epsilon}.
\eeq

The effective theory amplitude is obtained taking the limit 
$E\rightarrow\infty$ in the integrand:
\ber\label{pinch2}
C_E(r=0)&\sim&\int d^2 k_T \int d k_0 d k_z~
\frac{1}{k_0-k_z+i\epsilon}~\frac{1}{k_0-k_z-i\epsilon}
\nonumber\\
&\sim&\int d^2 k_T~\int d k_+ \int d k_-
\frac{1}{k_-+i\epsilon}~\frac{1}{k_--i\epsilon}
\eer
where $r$ is the meson residual momentum, which we set to zero.

The integrations over the transverse momentum and over $k_+$ give
the cubic ultraviolet divergence found before.
The integral over $k_-$ involves instead a pinch singularity
due to the infinitesimally close poles at $k_-=\pm i\epsilon$.
We note that pinching occurs in the whole transverse momentum space,
while it occurs only for $k_T\rightarrow 0$ in the full
theory.
Integrating over $k_-$ we pick up a $1/\epsilon$, i.e. a 
linearly divergent contribution:
\beq
C_E\sim \frac{1}{\epsilon}
\eeq
Thus the infrared behaviour of the full theory is not reproduced by the
$LEET$: 
a logarithmic infrared singularity in the full theory is in fact converted
into a linear (i.e. much stronger) singularity in the effective theory.

The conclusion is that the $LEET$ cannot be consistently applied to
exclusive processes.
Radiative corrections (powers in $\alpha_S$)
and power suppressed corrections (powers in $1/E$) do not
make the situation better.
As a typical example of a radiative correction, let us consider the
vertex correction of order $\alpha_S$ to the meson propagator, which
gives an integral of the form
\beq\label{fiveprop}
V~\sim~\int dk_+dk_-d^2k_T~dl_+dl_-d^2 l_T~ 
\frac{1}{k_-+i\epsilon}~\frac{1}{k_--i\epsilon}~
\frac{1}{(k-l)^2}~
\frac{1}{l_-+i\epsilon}~\frac{1}{l_--i\epsilon}~~~~
\eeq
The diagram is affected by pinch singularities
of higher order with respect to the tree diagram
\footnote{As we have shown before, diagram (\ref{fiveprop}) 
vanishes together with all the radiative corrections after making the
trace over color, but here we look only at 
the singular structure of the integrand.}.

Corrections in $1/E$ are more singular than the lowest order term
because they give rise to expressions of the form
\beq
\frac{1}{(k_-+i\epsilon)^2}~\frac{\vec{k_T}^2}{2E}~
\frac{1}{k_--i\epsilon},~~~~
\frac{1}{k_-+i\epsilon}~\frac{1}{(k_- -i\epsilon)^2}
\frac{\vec{k_T}^2}{2E},~\ldots
\eeq
These integrands have pinch singularities of higher order with 
respect to the tree diagram.
Therefore the `disease' is not even cured by corrections in $1/E$,
which actually made the situation worse. 

\subsection{Analogies with the $HQET$}
\label{secexclpert2}

There is a close analogy with the Heavy Quark Effective Theory
$(HQET)$ when the latter is applied to describe quarkonium states
(for example the $b\overline{b}$-spectroscopy).
The quarkonium propagator $Y$ (the analog of the light meson
propagator in eq.(\ref{dibase})) 
involves a tree level contribution in the full theory
\ber\label{fullmassivo}
Y_F&\sim& 
\int d^4 k \int
~\frac{1}{(P+k)^2-M^2+i\epsilon}~\frac{1}{(-P+k)^2-M^2+i\epsilon}~
\nonumber\\
&\sim&
\int d^3 k \int
dk_0~\frac{1}{k_0+k^2/(2M)+i\epsilon}~\frac{1}{k_0-k^2/(2M)-i\epsilon}
\eer 
where we have taken an external momentum $2P$ with $P=(M,\vec{0})$.
In the complex $k_0$-plane there are four poles at
\ber
k_0&=&\pm\sqrt{k^2+M^2}-M\mp i\epsilon,  
\nonumber\\
k_0&=&\pm\sqrt{k^2+M^2}+M\mp i\epsilon,  
\eer
There is a pinching of the poles at $k_0=\pm\sqrt{k^2+M^2}\mp M\mp
i\epsilon$ when $k=$ $\mid \vec{k}\mid\rightarrow 0$.  
Integrating over $k_0$ we get
\beq
Y_F\sim \int \frac {d^3 k}  {\vec{k}^2/M+O(\vec{k}^4)+i\epsilon}, 
\eeq
i.e. an infrared finite amplitude \footnote{In the massless case we found
instead a logarithmic term related to the collinear singularity.}.

The $HQET$ amplitude is obtained taking the limit $M\rightarrow\infty$
in the integrand:
\beq
Y_E(r^{\mu}=0)~\sim~\int d^3 k \int
dk_0~\frac{1}{k_0+i\epsilon}~\frac{1}{k_0-i\epsilon}
\eeq 
In the $k_0$-plane there are poles at $k_0=\pm i\epsilon$, so that
there is a pinch singularity for every value of $\vec{k}$.
The integral in $k_0$ is consequently divergent as $1/\epsilon$,
contrary to the full theory amplitude which is finite for
$\epsilon\rightarrow 0$.
The infrared behaviour of the full theory is consequently not
correctly reproduced by the $HQET$.

Corrections of order $1/M$ involve higher order pinch singularities,
because they generate integrands of the form
\beq
\frac{1}{(k_0+i\epsilon)^2}~\frac{\vec{k}^2}{2M}~
\frac{1}{k_0-i\epsilon},~~~~
\frac{1}{k_0+i\epsilon}~\frac{1}{(k_0-i\epsilon)^2}\frac{\vec{k}^2}{2M},~
\ldots
\eeq

In the case of mesons composed of a heavy and a light quark, 
(such as for example a $B$-meson), the heavy quark can be consistently
treated in the $HQET$ because
the light quark $q$ is described by the {\it full} theory and 
carries all the dynamics.
The propagator of $q$ can be written in a
background gauge field, because of covariance, as 
\cite{polyakov}
\beq
iS_F(x\mid 0;A_{\mu})~=~\sum_C f[C]~P~e^{i\int_C A_{\mu}
dx^{\mu}}  
\eeq
where $f[C]$ is a functional and its argument $C$ is a any path connecting
points $0$ and $x$.
Unlike the $LEET$, $iS_F$ is not concentrated on a single path
(the classical one), but it involves many paths with the
associated (many) P-lines.
These properties of $iS_F$ make the $B$-meson propagator 
non-singular and  non-trivial.

The meson propagator is given in the full theory in lowest order 
perturbation theory by
\ber
C_F &\sim& \int d^4 k~
\frac{1}{k^2+i\epsilon}~\frac{1}{(P+k)^2-M^2+i\epsilon}~
\nonumber\\
&\sim&
\int d^3 k
dk_0~\frac{1}{k_0-E_k+i\epsilon}~\frac{1}{k_0+E_k-i\epsilon}~
\frac{1}{k_0+k^2/(2M)+i\epsilon}
\eer
where $E_k=\mid \vec{k}\mid$ since we have taken $q$ massless for
simplicity.  
There are poles in the $k_0$-plane at
\ber
k_0&=&\pm\sqrt{k^2+M^2}-M\mp i\epsilon,  
\nonumber\\
k_0&=&\pm E_k\mp i\epsilon.  
\eer
There is a pinching of the poles at $k_0=\sqrt{k^2+M^2}-M-i\epsilon$
and $k_0=-E_k+i\epsilon$. The distance between these poles is
\beq
d=E_k+\sqrt{k^2+M^2}-M\simeq k+O(k^2),
\eeq  
i.e. it is linear in $k$ for $k\ll M$.

The $HQET$ amplitude is given by:
\beq
C_{HQET}~\sim~
\int d^3 k
dk_0~\frac{1}{k_0-E_k+i\epsilon}~\frac{1}{k_0+E_k-i\epsilon}~
\frac{1}{k_0+i\epsilon}.
\eeq 
The poles of the heavy quark propagator are replaced 
by a single pole at $k_0=-i\epsilon$,
so there is a pinching between the latter pole and the one at 
$-E_k+i\epsilon$. The distance between them is
\beq
d=k, 
\eeq
so it goes to zero with $k$ as in the full theory.
This implies that the infrared behaviour of the correlator is the same
in the full theory and in the $HQET$.

The above analysis confirms the validity of the $HQET$ approach to the
description of heavy-light systems.

\section{Factorization in Seminclusive Processes}
\label{secfac}

In this section we show that the $LEET$ does describe seminclusive
processes and can be used to prove approximate factorization
(in a sense specified below) in the process
\beq\label{ancora}
B~\rightarrow~D^{(*)}~+~jet
\eeq
in the limit
\beq\label{hardlim}
m_b-m_c~\rightarrow~\infty.
\eeq

If we treat the jet as a massless system, its energy is given 
in the $COM$ frame by 
\beq
E_{jet}~=~\frac{m_B^2-m_D^2}{2m_B}.
\eeq
The limit in which factorization holds is  
$E_{jet}~\rightarrow~\infty,$
which is implied by the limit (\ref{hardlim}), because
\footnote{Note in particular that for factorization to hold
it is not necessary to send the charm
mass to infinity, but only to send the beauty
mass to infinity (the latter acts as an energy resevoir).}:
\beq
E_{jet}~>~\frac{m_B-m_D}{2}~\simeq~\frac{m_b-m_c}{2}.
\eeq

To prove factorization, we just have to reinterpret the results
of sec.\ref{secexcl}.
Replacing the $up$ and $down$ quarks with $LEET$
quarks with the same velocity $n$ in eq.(\ref{factoriz}), 
their color interactions
with the heavy system disappear and factorization comes out. 
Actually, the problem is that we have a stronger property than
factorization: the light quark and the light antiquark do not have any
color interaction also with each other. However,
in the seminclusive process the latter property does not cause any
problem, because the $u$ and $d$ quarks have to be considered
hard partons, i.e. short distance excitations, unrelated to
exclusive dynamics.
 
To summarize: in the seminclusive case, the  
independence on the gauge field $A_{\mu}$ of the light quark
trace in eq.(\ref{factoriz}) {\it does mean} factorization.

In sec.\ref{secexclpert} we found that the correlators of the $LEET$ are
affected by strong pinch singularities in the perturbative expansion.
In a seminclusive approach these singularities do not occur anymore
because jets are defined integrating over energy and momentum
intervals. 
Pinch singularities disappear from the correlations after integrating 
in the region of phase space specified by the jet definition.
For the 2-point function, for example, eq.(\ref{pinch2})
is replaced by an integral of the form:
\ber
&&\int dp_-~f(p_-)~\int_{-\infty}^{\infty} d k_-~\frac{1}{k_- +p_- +
i\epsilon}~
\frac{1}{k_- -i\epsilon}
\nonumber\\
&=&2\pi i\int dp_-~f(p_-)~\frac{1}{p_-+i\epsilon}~
=~(definite~and~finite).
\eer
where $f(p_-)$ is a shape function, whose form
depends on the specific definition of the jet.

To summarize, we proved in a non-perturbative way that
a pair of hard quarks moving along
the same classical trajectory do not have any color interaction.
This implies, in particular, that they do not produce any radiation
field.
The suppression of the radiation field from a pair of hard
partons in a color singlet state at small angular separation
is a well know phenomenon in perturbative $QCD$.
It is a particular case of the so called `color coherence'
\cite{dks}. The latter is an interference effect
according to which a quark and an antiquark at
an angular separation $\delta$ in a color singlet state
generate a radiation field restricted to a cone of
width $\delta$. Outside the cone, complete destructive
interference takes place and no radiation is emitted. 
In this respect the decay (\ref{ancora}) 
is analogous to the high-energy annihilation:
\beq
e^+e^-~\rightarrow~q+\overline{q}+\gamma
\eeq
with the photon 
energy $E_{\gamma}$ close to its kinematical endpoint,
\beq
E_{\gamma}~\sim~\frac{\sqrt{s}}{2}~~~~~~~(COM~frame).
\eeq
The quark and the antiquark are in a color singlet state
and are emitted at a small angle $\delta\ll 1$
with the $\gamma$ recoiling in the opposite direction.
According to color coherence, secondary soft partons are emitted
at angles 
\beq
\theta~<~\delta
\eeq 
with respect to the quark and antiquark line of flight. In the limit 
\beq
\delta~\rightarrow~0^+,
\eeq
no radiation field is emitted. 
Another process analogous to the decay (\ref{ancora}) 
is the radiative decay of the $Y$,
\beq
Y~\rightarrow~g~+~g~+~\gamma
\eeq
close to the endpoint of the photon spectrum,
\beq\label{endp}
E_{\gamma}~\sim~m_b.
\eeq
The gluon pair is emitted in a color singlet state at a small
angular separation $\delta\ll 1$.
As in the decay (\ref{ancora}), one can prove in a non-perturbative way
that there are no color interactions of the gluon pair
in the limit $\delta\rightarrow 0$ at leading order in $1/N_C$.
The vanishing of color interactions at the endpoint (\ref{endp})
implies, in particular, that there is no the typical
Sudakov suppression of the differential rate $d\Gamma/dE_{\gamma}$
close to the endpoint
\footnote{We wish to thank G. Martinelli for having explained
this point to us.}, as we have for example in the
semileptonic $b\rightarrow u$ decay \cite{corbo}
\beq
b~\rightarrow~u~+~e~+~\nu_e.
\eeq

\subsection{Corrections to Factorization}

There are corrections to the factorization in the seminclusive
process (\ref{ancora}) related to the fact that the jet formed by the
light $ud$ pair is not infinitely narrow. 
A quantitative discussion requires a detailed definition of a jet,
such as for example the Sterman-Weinberg one \cite{sterwein},
which is beyond the scope of this paper.
We present only a qualitative discussion.
Let us assume a jet angular width 
\beq
2~\delta~\ll~1.
\eeq
This implies that the light pair can be emitted 
with a relative angle up to
$2~\delta$, so that the light-like vectors $n$ and $n'$ of the
quark and the antiquark respectively, can be written 
up to first order as
\beq
n~=~(1;\delta,0,1),~~~~~~~~~~n'~=~(1;-\delta,0,1).
\eeq
We have taken the relative motion 
of the pair in the $x$ direction and $n^2=n'^2=\delta^2\sim0$.

\subsubsection{Jet with Finite Angular Width} 

As usual, let us begin by considering the simplest case, a 
3-point correlation function representing the creation
of a pair of $LEET$ quarks moving along $n$ and $n'$
\beq
J_3(x,x')~=~\langle 0\mid T~B(x,x')~L^{\dagger}(0) \mid 0 \rangle
\eeq
where $L(y)$ is a local operator which annihilates a pair
of effective quarks with velocities $n$ and $n'$:
\beq
L(y)~=~\overline{Q}_{n'}(y)\Gamma Q_n(y),
\eeq
while $B(x,x')$ is a bilocal operator which,
by covariance, may be written as
\beq
B(x,x')~=~\overline{Q}_{n'}(x')~\Gamma~ 
P\exp\Bigg[ig\int_{x}^{x'} A_{\mu} dx^{\mu}\Bigg]~Q_{n}(x)
\eeq
There is an ambiguity (of non-perturbative kind) in the choice of the path
connecting point $x$ with $x'$. We consider small angles of emission of
the quarks and we simply take as the path the segment joining $x$
with $x'$.

This correlation function gives a nonperturbative
representation of a jet, so we may
call it the jet correlator.
The functional representation of the jet correlator is
\beq
\langle 0\mid T~B(x,x')~L^{\dagger}(0) \mid 0 \rangle_A~
=~-Tr\Big[iS_n(x\mid 0)\tilde{\Gamma} iS_{n'}(0\mid x')~
\Gamma~P(x'\mid x)\Big]~~~~~
\eeq
Inserting the expressions for the $LEET$ propagators, we derive:
\ber\label{chilosa}
J_3(x,x')&=&-\theta(t)~\theta(t')~
\frac{ \delta^{(3)}(\vec{x}-\vec{u}t) }{n_0}~
\frac{ \delta^{(3)}(\vec{x'}-\vec{u}'t') }{n_0'}~
\nonumber\\
&&N\int [dA]~e^{iS_{eff}[A]}
Tr\Big[~P(x\mid 0)~P(0\mid x')~P(x'\mid x)\Big]~~~~
\eer
where a more compact notation $P(y\mid x)$ has been assumed
for a P-line joining $x$ with $y$.

From eq.(\ref{chilosa}) we see that
the estimate of the jet correlator requires a non-perturbative
computation of a Wilson loop on a triangular path, having
as vertices the points 
\beq
x~=~0,~~x=n\tau,~~x'=n'\tau',
\eeq
with some selected value for $\tau$ and $\tau'$.

\subsubsection{Finite Jet Width in Non-Leptonic Decay}

The seminclusive process (\ref{ancora}) 
is related to the following 5-point function
\beq
J_5(z,z',y,x)~=~\langle 0\mid T~B(z,z')~O_D(y)~\cal{H}_W(0)~ 
O_B^{\dagger}(x) \mid 0 \rangle
\eeq
where the weak hamiltonian contains two $LEET$ fields
with different velocities $n$ and $n'$, so that the operators 
$O_1$ and $O_8$ in eq.(\ref{effham})  are of the form
\beq
O_i(x)~=~\overline{Q}_{n}(x)~\gamma_{\mu L}\xi_i~Q_{n'}(x)~
\overline{c}(x)~\gamma^{\mu}_L\xi_i~b(x)
\eeq
We have:
\ber
&&\langle 0\mid T~B(z,z')~O_D(y)~\cal{H}_W(0)~ 
O_B^{\dagger}(x) \mid 0 \rangle_{A}
\nonumber\\
&=&\sum_{i=1,8}~
C_i~Tr\Big[~iS_c(y\mid 0)~\gamma_{\mu L}\xi_i~iS_b(0\mid x)~
\tilde{\Gamma}_B~
iS_{s}(x\mid y)~\Gamma_D~\Big]
\nonumber\\
&&~~~~ Tr\Big[iS_n(z\mid 0)~\gamma^{\mu}_L\xi_i~
iS_{n'}(0\mid z')~\Gamma~P(z'\mid z)~\Big]
\nonumber\\
&=&\sum_{i=1,8}~
C_i~Tr\Big[~iS_c(y\mid 0)~\gamma_{\mu L}\xi_i~iS_b(0\mid x)~
\tilde{\Gamma}_B~
iS_{s}(x\mid y)~\Gamma_D~\Big]
\nonumber\\
 &&~~~~\theta(t_z)~\theta(t_z')~
\frac{ \delta^{(3)}(\vec{z}-\vec{u}t_z) }{n_0}~
\frac{ \delta^{(3)}(\vec{z'}-\vec{u}'t_z') }{n_0'}~ Tr\Bigg[~
\frac{\hat{n}}{2}\gamma^{\mu}_L~\frac{\hat{n'}}{2}\Gamma~\Bigg]~
\nonumber\\
&&~~~~~~~ Tr\Big[~P(z\mid 0)~\xi_i~P(0\mid z')~P(z'\mid z)~\Big].
\eer
Unlike the previous case in which $n'=n$, the P-lines do not cancel
each other any more and there is a dependence on the gauge field: 
gauge dynamics is not trivial, 
factorization does not hold anymore
and both the operators $O_1$ and $O_8$ give a
non-vanishing contribution to the correlator. 
The corrections to factorization are related to matrix elements of the
following form:
\beq
\langle D^{(*)} \mid T_i~J_{\mu,i}^{b\rightarrow c}(0) 
\mid B\rangle
\eeq
where $J_{\mu,i}^{b\rightarrow c}(x)=
\overline{c}(x)\gamma_{\mu}^L \xi_i b(x)$ and
$T_i$ is the following gauge invariant/covariant operator:
\beq
T_i~=~Tr\Big[~P(z\mid 0)~\xi_i~P(0\mid z')~P(z'\mid z)~\Big],
\eeq
i.e. it is the trace of a Wilson loop on the thin triangle considered
in the previous section, with the insertion of $\xi_i=1,t_a$ at the 
vertex with the small angle $2\delta\ll 1$.

According to this line of reasoning, we believe that factorization
should be strongly violated in multiple jet production, 
i.e. in processes like
\beq\label{2jets}
B~\rightarrow~D^{(*)}~+~2~jets.
\eeq
That is because,
according to the picture of the process given in the introduction,
in this process there is a large color dipole field of the light
quarks, strongly interacting with the $B$ and $D^{(*)}$ mesons.

\section{A New Effective Theory}
\label{secnew}

We can remedy to the inadequacy of the $LEET$ to describe
exclusive processes by including 
the leading kinetic correction (see eq.(\ref{drl2})) into the propagator:
\beq\label{nft}
iS_F(k)~=~\frac{\hat{n}}{2}~
\frac{i}{n\cdot k-{\vec{k}_T}^2/2E~+i\epsilon}
\eeq
where we may take $n^{\mu}=(1;0,0,1),~k_T^{\mu}=(0;\vec{k}_T,0)$
\footnote{
We can give a Lorentz invariant representation of the
transverse momentum $k_T$ with the Sudakov basis. 
If we define a second light-like vector $\eta$ such that 
$n\cdot \eta=2$ (in the usual frame, $\eta=(1;0,0,-1)$),
we have $n\cdot k_T=\eta\cdot k_T=0$ and
$k_T^2=-\vec{k}_T^2=k^2-n\cdot k~\eta\cdot k$.}.
In the derivation of eq.(\ref{nft}) we followed an idea of `minimal 
correction' of the $LEET$ pathologies; we have neglected for example
the term $\hat{k}$ in the numerator of eq.(\ref{nft}),
so that the spin structure is factorized. 

We may call this new effective theory `Modified Large
Energy Effective Theory', $\overline{LEET}$ for short.
Note that, unlike the $LEET$ case, the hard scale $E$
is still present in the theory, i.e. it cannot be completely
removed.

The problem of pinch singularities discussed in sec.\ref{secexclpert}
is solved replacing $LEET$ propagators with $\overline{LEET}$ 
propagators: the meson propagator in the $\overline{LEET}$ 
has the form (\ref{approx}) so that pinch singularities 
occur only for $\vec{k}_T=0$ instead of in the whole transverse
momentum space. 
The $\overline{LEET}$ amplitude 
coincides with the full theory 
amplitude (\ref{nonapprox}) for $k_+\ll E$ and has the same infrared
behaviour.

Eq.(\ref{nft}) still defines an effective theory,
even though much more complicated than the $LEET$:
the propagator contains the 4-velocity $n$,  
is forward in time, so that antiparticles
are removed and the vacuum is consequently 
trivial. This is just what we expect from an effective theory
describing hard partons, in which  
particle-antiparticle pairs have a threshold energy
of order $2E$ and cannot be excited with soft interactions.
Furthermore, their virtual effect is expected to be small on the basis
of the decoupling theorem \cite{apc}.

The propagator is given as a function of time and spatial momentum
by
\beq\label{kappat}
iS_F(t,\vec{k})~=~\frac{\hat{n}}{2}~\theta(t)~
\exp\Bigg[-ik_Zt-i\frac{k_T^2~t}{2E}~\Bigg]
\eeq
and in configuration space by 
\ber\label{spatim}
iS_F(t,\vec{x})&=&\int
\frac{d^3k}{(2\pi)^3}~e^{i\vec{k}\cdot\vec{x}}~
               iS_F(t,\vec{k})
\nonumber\\
&=&\frac{\hat{n}}{2}~
\theta(t)~\delta(z-t)~\frac{E}{2\pi it}~e^{iE b^2/(2t)}
\nonumber\\
&=&iS_F^{LEET}(x)~\frac{E}{2\pi i t}e^{i E b^2/(2t)}
\eer
where $b=\mid\vec{x}_T\mid$ is the impact parameter.
The effect of the transverse momentum term is factorized
and produces a diffusion in the impact parameter space: 
the factor in the last line of eq.(\ref{spatim}) represents 
a gaussian process after analytic continuation
$t_M=-it_E$.
At large times, we have:
\beq
S(t,\vec{x})~\simeq~\frac{\hat{n}}{2}~
\theta(t)~\delta(z-t)~\frac{E}{2\pi it}.
\eeq
We see that there is a diffusion normal to the classical particle
trajectory $z=t$ produced by transverse momentum fluctuations,
which is instead absent in the $LEET$.
The amplitude for the particle to remain into the classical
trajectory decays like $1/t$, so the probability decays like
$1/t^2$.

The lagrangian of the $\overline{LEET}$, omitting the spin
dependence, is 
\beq
\cal{L}(x)~=~Q^{\dagger}(x)\Bigg[~in\cdot D+\frac{{D_T}^2}{2E}~\Bigg]Q(x).
\eeq

We believe that the $\overline{LEET}$ is the correct effective
theory for massless particles as long as exclusive processes
are concerned. 

Let us make a general observation. As we have shown in detail
in sec.\ref{secexcl}, $1/E$ corrections cannot be considered
perturbations in exclusive processes (i.e. local operator insertions),
even though they are related to higher dimension operator:
they must be kept in the unperturbed
propagator. For example, expanding the propagators (\ref{nft}) in
powers of $1/E$ in the meson propagator (\ref{dibase}) 
is not a `legal' operation. Here we have a counter-example
of the general validity of Wilson's Operator Product Expansion
\cite{wil}: the kinetic operator, a higher dimensional
operator, cannot be considered `irrelevant' because of
infrared effects.

It is hard to reach a conclusion about factorization in the
exclusive decay (\ref{nld}) on the basis of simple analytical
computations with the $\overline{LEET}$.
That is because of diffusion in the impact parameter space, according
to which light quark dynamics is represented by a superposition
of Wilson loops on strips with a width of order $b$.
A non-perturbative technique is needed:
the $\overline{LEET}$ lagrangian can be discretized 
on an euclidean lattice and its dynamics can be computed
numerically with lattice $QCD$. 

We think that the $\overline{LEET}$ can describe also some 
perturbative $QCD$ effects in which transverse momentum 
dynamics is important, and cannot be neglected 
altogether as in the $LEET$, such as for example initial
state interactions in Drell-Yan processes \cite{drellyan}.
We argue that the absence of antiparticle excitations
is related to the analogous property of the vacuum state
in the infinite momentum frame.

\subsection{Analogies with Non Relativistic $QCD$}

The $\overline{LEET}$ is the analog of Non Relativistic $QCD$
$(NRQCD)$ for massive quarks,
which {\it must} be used in place of the $HQET$ to 
describe quarkonium dynamics.
The propagator of the $NRQCD$ encodes the non-relativistic
energy-momentum relation and is given by
\beq\label{nrel}
iS_F(k)~=~\frac{1+\gamma_0}{2}~\frac{i}{k_0-\vec{k}^2/2M+i\epsilon}.
\eeq
The basic approximations are:
\begin{enumerate}
\item[$(i)$]
the spin fluctuation $\hat{k}$ is neglected;
\item[$(ii)$]
the antiparticle pole is eliminated, so that
\beq
k^2~\rightarrow~-\vec{k}^2.
\eeq
\end{enumerate}
The main point is that the term $\vec{k}^2/(2M)$ is
kept in the denominator, i.e. no
expansion in $1/M$ is done as in the $HQET$.
This means that a partial resummation of $1/M$ corrections
is performed.

The problem of pinch singularities in quarkonium propagators 
discussed in sec.\ref{secexclpert2} is eliminated
replacing $HQET$ propagators with $NRQCD$ propagators \cite{lepage}. 
The quarkonium propagator in $NRQCD$ has the form:
\beq
Y_{NRQCD} \sim 
\int d^3 k \int
dk_0~\frac{1}{k_0-\vec{k}^2/(2M)+i\epsilon}~
\frac{1}{k_0+\vec{k}^2/(2M)-i\epsilon}~
\eeq
The pinching of the poles occurs only in the point $\vec{k}=0$
instead of in the whole 3-momentum space as with $HQET$ propagators.
Performing the integration over $k_0$ we derive:
\beq
Y_{NRQCD}~\sim~\int d^3 k~\frac{1}{\vec{k}^2/M-i\epsilon}
\eeq
which is infrared finite as the full theory correlator.
Therefore the $NRQCD$ propagator
has the same infrared behaviour of the full theory
propagator (eq.(\ref{fullmassivo})).
\section{Conclusions}
\label{secconcl}

The main conclusion of our analysis is that the Large Energy
Effective Theory can be used to prove factorization
in the {\it semi inclusive} process
\beq
B~\rightarrow~D^{(*)}~+~jet
\eeq
in the limit 
\beq
m_b-m_c~\rightarrow~\infty.
\eeq
We tried our best to convince the reader that factorization in
seminclusive decays can be proved pretty rigorously 
in a non-perturbative way with the theory of Wilson loops.
The idea is that,
in the limit of a very narrow jet, the Wilson loop collapses
into a rectangle with two infinitesimal edges, which 
does not depend on the gauge field any more and
can be either one or zero, depending on the operators
acting at the vertices.

Factorization is exact in the limit of an infinitely narrow jet,
and corrections related to a finite angular width $\delta\ll 1$ 
can be written in a gauge invariant way in terms of Wilson loops
on triangular paths.
The evaluation of these corrections requires a non-perturbative
technique, such as for example lattice $QCD$.
Following the same line of reasoning, 
we argue that factorization should be strongly violated in multiple
jet production, i.e. in processes like for example
\beq
B~\rightarrow~D^{(*)}~+~2~jets.
\eeq

On the other hand, we believe that factorization cannot be
rigorously proved in {\it exclusive} non-leptonic decays,
such as for example
\beq
B~\rightarrow~D^{(*)}+\pi(\rho).
\eeq
That is because the theoretical tool,
the Large Energy Effective Theory, is intrinsically incapable to describe
bound state effects and exclusive hadron dynamics.
This is due to neglecting  transverse momentum dynamics.
We have introduced a new effective theory for massless particles,
the Modified Large Energy Effective Theory,
which takes into account transverse momentum dynamics, and
which is the right framework for studying exclusive processes.
The latter theory is however more complicated so
we did not succeed in deriving any conclusion about factorization
on the basis of analytical computations with it.

Finally, we have also shown that the Large Energy Effective Theory is 
a consistent theory once seminclusive observables are
considered instead of completely exclusive ones.

$~~$

\centerline{\bf Acknowledgements}

$~~$

We wish to thank G. Martinelli and C. Sachrajda for discussions.

$~~$

\vfill

\end{document}